\documentclass[10.5pt]{article}

\def\mytitle#1{\setcounter{equation}{0}
\setcounter{footnote}{0}
\begin{flushleft}\Large\textbf{#1}\end{flushleft}
\vspace{0.25cm}}
\def\myname#1{\leftline{{\large #1}}\vspace{-0.13cm}}
\def\myplace#1#2{\small\begin{flushleft}\textit{#1}\\
\texttt{#2}\end{flushleft}}

\def\myclassification#1{\small\noindent
Pacs no :
       #1\vspace{0.5cm}}
\usepackage{graphicx}% Include figure files
\usepackage{amsmath, amssymb, graphics}
\newcommand{\mathsym}[1]{{}}

\begin{document}

\mytitle{\Large Dynamical System Analysis of Modified Chaplygin
Gas in Einstein-Aether Gravity}

\vskip0.2cm \myname{Chayan
Ranjit\footnote{chayanranjit@gmail.com}} \myplace{Department of
Mathematics, Seacom Engineering College, Howrah-711 302, India.}{}
\vskip0.2cm

\vskip0.2cm \myname{Prabir Rudra\footnote{prudra.math@gmail.com}}
\myplace{Department of Mathematics, Indian Institute of Engineering Science and Technology, Shibpur, Howrah-711 103, India.\\
Department of Mathematics, Pailan College of Management and
Technology, Bengal Pailan Park, Kolkata-700 104, India.}{}

\vskip0.2cm \myname{Sujata
Kundu\footnote{sujatakundu10@gmail.com}} \myplace{Department of
Information Technology, Narula Institute of Technology,
Kolkata-700109,India}{} \vskip0.2cm

\begin{abstract}
In this work we investigate the background dynamics when dark
energy is coupled to dark matter with a suitable interaction in
the universe described by Einstein-Aether gravity. Dark energy in
the form of Modified Chaplygin gas is considered. A suitable
interaction between dark energy and dark matter is considered in
order to at least alleviate (if not solve) the cosmic coincidence
problem. The dynamical system of equations is solved numerically
and a stable scaling solution is obtained. A significant attempt
towards the solution of the cosmic coincidence problem is taken.
The statefinder parameters are also calculated to classify the
dark energy models. Graphs and phase diagrams are drawn to study
the variations of these parameters. It is also seen that the
background dynamics of modified Chaplygin gas in Einstein-Aether
gravity is completely consistent with the notion of an accelerated
expansion in the late universe. Finally, it has been shown that
the universe follows the power law form of expansion around the
critical point.
\end{abstract}

\myclassification{04.50.Kd, 95.36.+x, 98.80.Cq, 98.80.-k}

\section{\normalsize\bf{Introduction}}

At the turn of the last century observations from Ia Supernova and
Cosmic Microwave Background (CMB)radiation confirmed that our
universe is suffering from an accelerated expansion
\cite{Perlmutter,Riess,Riess1,Bennet,Sperge}, but the physical
origin of this acceleration is yet to be known. The standard
explanation invokes an unknown ``dark energy'' component which has
the property that positive energy density and negative pressure.
Observations indicate that dark energy occupies about 70\% of the
total energy of the universe, and the contribution of dark matter
is $\sim$ 26\%. This accelerated expansion of the universe has
also been strongly confirmed by some other independent experiments
like Sloan Digital Sky Survey (SDSS) \cite{Adel}, Baryonic
Acoustic Oscillation (BAO) \cite{Eisenstein}, WMAP data analysis
\cite{Briddle,Spergel} etc. Over the past decade there have been
many theoretical models for mimicking the dark energy behaviors,
such as the simplest (just) cosmological constant in which the
equation of state is independent of the cosmic time and which can
fit the observations well. This model is the so-called
$\Lambda$CDM, containing a mixture of cosmological constant
$\Lambda$ and cold dark matter (CDM). However, two problems arise
from this scenario, namely ``fine-tuning'' and the ``cosmic
coincidence'' problems. In order to solve these two problems, many
dynamical dark energy models were suggested, whose equation of
state evolves with cosmic time. The scalar field or quintessence
\cite{Peebles,Cald} is one of the most favored candidate of dark
energy which produce sufficient negative pressure to drive
acceleration. In order to alleviate the cosmological-constant
problems and explain the acceleration expansion, many dynamical
dark energy models have been proposed, such as K-essence, Tachyon,
Phantom, quintom, Chaplygin gas model, etc
\cite{Arme,Sen,Cald1,Feng,Kamen}. Also the interacting dark energy
models including Modified Chaplygin gas \cite{Debnath},
holographic dark energy model \cite{Cohen}, and braneworld model
\cite{Sahni} have been proposed. The equation of state of Modified
Chaplygin gas is given by,

\begin{equation}
p=A\rho-\frac{B}{\rho^\alpha},
\end{equation}
where $p$ and $\rho$ are respectively the pressure and energy
density and $0\leq\alpha\leq1$, $A$ and $B$ are positive
constants. In Einstein's gravity, the modified Chaplygin gas
\cite{Debnath} best fits with the 3 year WMAP and the SDSS data
with the choice of parameters $A =0.085$ and $\alpha = 1.724$
\cite{Lu} which are improved constraints than the previous
ones $-0.35 < A < 0.025$ \cite{Jun}.\\

Another possibility is that general relativity is only accurate on
small scales and has to be modified on cosmological distances. One
of these is modified gravity theories. In this case cosmic
acceleration would arise not from dark energy as a substance but
from the geometry of space-time i.e. from the dynamics of modified
gravity. Modified gravity constitutes an interesting dynamical
alternative to $\Lambda$CDM cosmology in that it is also able to
describe the current cosmic acceleration. The simplest modified
gravity is DGP brane-world model \cite{Dvali}. The other
alternative approach dealing with the acceleration problem of the
Universe is changing the gravity law through the modification of
action of gravity by means of using $f(R)$ gravity \cite{An,Noj0}
instead of the Einstein-Hilbert action. Some of these models, such
as $1/R$ and logarithmic models, provide an acceleration for the
Universe at the present time \cite{clif}. Other modified gravity
includes $f(T)$ gravity, $f(G)$ gravity, Gauss-Bonnet gravity,
Horava-Lifshitz gravity, Brans-Dicke gravity, etc
\cite{Yer,Noj,An1,Hora,Brans}. In recent times there have been a
lot of research on the background dynamics of different DE models
in modified
gravity theories in the quest of a standard model of cosmology \cite{Rudra1, Rudra2, Rudra3, Rudra4}.  \\

In the present work, we concentrate on the generalized
Einstein-Aether (EA) theories as proposed by Zlosnik et al
\cite{Zlos,Zlos1}, which is a generalization of the
Einstein-Aether theory developed by Jacobson et al
\cite{Jacob,Jacob1}. In recent years a lot of work has been done
in generalized Einstein-Aether theories
\cite{Gar,Linder,Barrow,Junt,Li,Gasp1,Gasp2,Deb}. In the
generalized Einstein-Aether theories by taking a special form of
the Lagrangian density of Aether field, the possibility of
Einstein-Aether theory as an alternative to dark energy model is
discussed in detail, that is, taking a special Aether field as a
dark energy candidate and it has been found the constraints from
observational data \cite{Meng1,Meng2}. Since modified gravity
theory may be treated as alternative to dark energy, so Meng et al
\cite{Meng1,Meng2} have not taken by hand any types of dark energy
in Einstein-Aether gravity and shown that the gravity may be
generates dark energy. Here if we exempt this assumption, so we
need to consider the dark energy from outside. So we assume the
FRW universe in Einstein-Aether gravity model filled with the dark
matter and the modified Chaplygin gas (MCG) type dark energy.

This paper is organized as follows: Section 2 comprises of the
general concepts of Einstein-Aether gravity. In Section 3, we do
an extensive study of the dynamical system. In section 4, a
detailed graphical analysis for the phase plane is presented.
Finally the paper ends with some concluding remarks in section 5.

\section{\normalsize\bf{Einstein-Aether Gravity Theory}}

In order to include Lorentz symmetry violating terms in
gravitation theories, apart from some noncommutative gravity
models, one may consider existence of preferred frames. This can
be achieved admitting a unit timelike vector field in addition to
the metric tensor of spacetime. Such a timelike vector implies a
preferred direction at each point of spacetime. Here the unit
timelike vector field is called the {\it Aether} and the theory
coupling the metric and unit timelike vector is called the {\it
Einstein-Aether} theory \cite{Jacob}. So Einstein-Aether theory is
the extension of general relativity (GR) that incorporates a
dynamical unit timelike vector field (i.e., Aether). In the last
decade there is an increasing interest in the Aether theory.\\

The action of the Einstein-Aether gravity theory with the normal
Einstein-Hilbert part action can be written in the form
\cite{Zlos,Meng1}
\begin{equation}
S=\int d^{4}x\sqrt{-g}\left[\frac{R}{16\pi G}+{\cal L}_{EA}+{\cal
L}_{m} \right]
\end{equation}

where ${\cal L}_{EA}$ is the vector field Lagrangian density while
${\cal L}_{m}$ denotes the Lagrangian density for all other matter
fields. The Lagrangian density for the vector part consists of
terms quadratic in the field \cite{Zlos,Meng1}:

\begin{equation}
{\cal L}_{EA}=\frac{M^{2}}{16\pi G}~F(K)+\frac{1}{16\pi
G}~\lambda(A^{a}A_{a}+1)~,
\end{equation}
\begin{equation}
K=M^{-2}{K^{ab}}_{cd}\nabla_{a}A^{c}\nabla_{b}A^{d}~,
\end{equation}
\begin{equation}
{K^{ab}}_{cd}=c_{1}g^{ab}g_{cd}+c_{2}\delta^{a}_{c}\delta^{b}_{d}+c_{3}\delta^{a}_{d}\delta^{b}_{c}
\end{equation}

where $c_{i}$ are dimensionless constants, $M$ is the coupling
constant which has the dimension of mass, $\lambda$ is a Lagrange
multiplier that enforces the unit constraint for the time-like
vector field, $A^{a}$ is a contravariant vector, $g_{ab}$ is
metric tensor and $F(K)$ ia an arbitrary function of $K$. From
(1), we get the field equations
\begin{equation}
G_{ab}=T_{ab}^{EA}+8\pi G T_{ab}^{m}~,
\end{equation}
\begin{equation}
\nabla_{a}\left(F'{J^{a}}_{b}\right)=2\lambda A_{b}
\end{equation}
where
\begin{equation}
F'=\frac{dF}{dK}~~and~~{J^{a}}_{b}=2{K^{ad}}_{bc}\nabla_{d}A^{c}
\end{equation}
Here $T_{ab}^{m}$ is the energy momentum tensor for matter field
and $T_{ab}^{EA}$ is the energy momentum tensor for the vector
field and they are respectively given as follows: \cite{Meng1}
\begin{equation}
T_{ab}^{m}=(\rho+p)u_{a}u_{b}+pg_{ab}
\end{equation}
where $\rho$ and $p$ are respectively the energy density and
pressure of matter and $u_{a}=(1,0,0,0)$ is the fluid 4-velocity
vector and
\begin{equation}
T_{ab}^{EA}=\frac{1}{2}~\nabla_{d}\left[
\left({J_{(a}}^{d}A_{b)}-{J^{d}}_{(a}A_{b)}-J_{(ab)}A^{d}
\right)F' \right]-Y_{(ab)}F'+\frac{1}{2}~g_{ab}M^{2}F+\lambda
A_{a}A_{b}
\end{equation}
with
\begin{equation}
Y_{ab}=-c_{1}\left[
(\nabla_{d}A_{a})(\nabla^{d}A_{b})-(\nabla_{a}A_{d})(\nabla_{b}A^{d})
\right]
\end{equation}
where the subscript $(ab)$ means symmetric with respect to the
indices involved and\textbf{ the vector $A^{a}$ is considered to
have time-like direction and satisfies $A^{a}A_{a}=-1$ to fix the
variation of the action with respect to $\lambda$. The normalized
vector field $A^{a}$ has components $(1,0,0,0)$ in the FRW
cosmology with homogeneous and isotropic universe filled
with perfect fluid.}\\

\section{Dynamical system analysis in Einstein-Aether gravity}

We consider the Friedmann-Robertson-Walker (FRW) metric of the
universe as
\begin{equation}
ds^{2}=-dt^{2}+a^{2}(t)\left[\frac{dr^{2}}{1-kr^{2}}+r^{2}\left(d\theta^{2}+sin^{2}\theta
d\phi^{2}\right) \right]
\end{equation}
where $k~(=0,\pm 1)$ is the curvature scalar and $a(t)$ is the
scale factor. From equations (3) and (4), we get
\begin{equation}
K=\frac{3\beta H^{2}}{M^{2}}
\end{equation}
where $\beta=c_{1}+3c_{2}+c_{3}$ is constant. From eq. (5), we get
the modified Friedmann equation for Einstein-Aether gravity as in
the following \cite{Zlos,Meng1}:
\begin{equation}
\beta\left(-F'+\frac{F}{2K}\right)H^{2}+\left(H^{2}+\frac{k}{a^{2}}\right)=\frac{8\pi
G}{3}~\rho
\end{equation}
and
\begin{equation}
\beta\frac{d}{dt}\left(HF'\right)+\left(-2\dot{H}+\frac{2k}{a^{2}}\right)=8\pi
G(\rho+p)
\end{equation}
where $H~(=\frac{\dot{a}}{a})$ is Hubble parameter. Now we see
that if the first expressions of L.H.S. of equations {\bf{(14)}}
and {\bf{(15)}} are zero, we get the usual field equations for
Einstein's gravity. So first expressions arise for Einstein-Aether
gravity. Also the conservation equation is given by
\begin{equation}
\dot{\rho}+3\frac{\dot{a}}{a}(\rho+p)=0
\end{equation}

Now, assume that the matter fluid is combination of dark matter
and modified Chaplygin gas type dark energy. So
$\rho=\rho_{m}+\rho_{ch}$ and $p=p_{m}+p_{ch}$, where $\rho_{m}$
and $p_{m}$ are respectively the energy density and pressure of
dark matter and $\rho_{ch}$ and $p_{ch}$ are respectively the
energy density and pressure of modified Chaplygin gas. Assume that
the dark matter follows the barotropic equation of state
$p_{m}=w_{m}\rho_{m}$, where $w_{m}$ is a constant. The equation
of state of modified Chaplygin gas (MCG) is given by
\cite{Debnath}
\begin{equation}
p_{ch}=A\rho_{ch}-\frac{B}{\rho_{ch}^{\alpha}}
\end{equation}

where $A>0$, $B>0$ and $0\le\alpha\le 1$.

As in the present problem the interaction between DE and
pressureless DM has been taken into account for interacting DE and
DM the energy balance equation will be
\begin{equation}
\dot{\rho}_{ch}+3H\left(1+\omega_{ch}\right)\rho_{ch}=-Q
\end{equation}
\begin{equation}
\dot{\rho}_m+3H\rho_m=Q
\end{equation}
where $Q=3bH\rho$ is the interaction term, $b$ is the coupling
parameter (or transfer strength) and $\rho=\rho_{ch}+\rho_m$ is
the total cosmic energy density which satisfies the energy
conservation equation $\dot{\rho}+3H\left(\rho+p\right)=0$
\cite{Guo1, del Campo1}.

Since we lack information about the fact, how does DE and DM
interact so we are not able to estimate the interaction term from
the first principles. However, the negativity of $Q$ immediately
implies the possibility of having negative DE in the early
universe which is overruled by the necessity of the second law of
thermodynamics to be held \cite{Alcaniz1}. Hence $Q$ must be
positive and small. From the observational data of 182 Gold type
Ia supernova samples, CMB data from the three year WMAP survey and
the baryonic acoustic oscillations from the Sloan Digital Sky
Survey, it is estimated that the coupling parameter between DM and
DE must be a small positive value (of the order of unity), which
satisfies the requirement for solving the cosmic coincidence
problem and the second law of thermodynamics \cite{Feng1}. Due to
the underlying interaction, the beginning of the accelerated
expansion is shifted to higher redshifts. The continuity equations
for dark energy and dark matter are given in equations (18) and
(19). Now we proceed to study the dynamical system.

\subsection{\normalsize\bf{Dynamical System Analysis}}

In this subsection we plan to analyze the dynamical system. For
that firstly we convert the physical parameters into some
dimensionless form, given by
\begin{equation}
x=\ln a, ~~ u=\frac{\rho_{ch}}{3H^2}, ~~ v=\frac{\rho_m}{3H^2},~~
y=\frac{a}{3H^{2}}
\end{equation}
where the present value of the scale factor $a_0=1$ is assumed.

\textbf{Following the works of Zlosnik et al and Zuntz et al
\cite{Zlosnik,Zuntz}, we assume the ansatz for $F$ as}
\begin{equation}
F(K)=f_{0}K^{n}~,~~~~~~~K>0
\end{equation}
\textbf{where cosmic acceleration is realized for $f_0>0$ and for
certain value of n $(n\leq 1)$ without the need of a source term
in the modified Einstein equations \cite{Zuntz}, here modelled
with the Chaplygin gas. This form is efficient enough to express a
wide range of behaviour. Following modified newtonian dynamics, if
we use the same branch of $F$ as it uses on small scales, it can
be shown that the background cosmology becomes inconsistent with
the late time cosmic acceleration \cite{Zuntz}. Hence we use a
reasonable form for $F$ given in eqn. (27), which works for the
accelerated regime ($|K|\gg 1$). Using this form of $F$, the
modified Friedmann equations become,}
\begin{equation}
\left[1+\epsilon\left(\frac{H}{M}\right)^{2(n-1)}\right]H^{2}=\frac{8\pi
G}{3}\rho
\end{equation}
\textbf{where
$\epsilon=\left(1-2n\right)f_{0}\left(-3\beta\right)^{n}/6$.
Moreover the expression for $f_{0}$ is obtained as,}
\begin{equation}
f_{0}=\frac{6\left(\Omega_{m}-1\right)}{\left(1-2n\right)\left(-3\beta\right)^{n}}\left(\frac{M}{H_{0}}\right)^{2(n-1)}
\end{equation}
\textbf{where $\Omega_{m}=8\pi G\rho_{0}/3H_{0}^{2}$ and $H_{0}$
is the present value of Hubble constant}.

\textbf{Let us consider some special cases of $n$. We see that if
$n=1/2$, the Friedmann equations are unchanged ($\epsilon=0$), and
invariably there is no effect on the background cosmology. If
$n=0$, the cosmological constant is recovered. When $n=1$, we have
$\epsilon=f_{0}\beta/2$ and there is a change of scale for the
Newton's constant by a factor of $1/(1+\epsilon)$ \cite{Caroll1}.
A schematic representation of the late time evolution of the
universe depending on the value of $n$ can be found in
\cite{Clifton2}. From the above discussion the most suitable
choice of $n$ for our present assignment is $n=1$. Moreover it
provides a bit of mathematical simplicity as well. This value of
$n$ completely lies in the accepted region which is $n\leq 1$. As
apparently conceived, this chosen value of $n$ does not actually
hamper the generality of the problem. In fact from our analysis we
have seen that a slight variation in the value of $n$ does not
bring about significant alterations to the final result.}

Using eqns. (13-19) into eqn.(20) we get the parameter gradients
as

\begin{equation}\label{8}
\frac{du}{dx}=\frac{H\dot{\rho_{ch}}-2\rho_{ch}\dot{H}}{3H^{4}}
\end{equation}

\begin{equation}\label{9}
\frac{dv}{dx}=\frac{H\dot{\rho_{m}}-2\rho_{m}\dot{H}}{3H^{4}}
\end{equation}
and

\begin{equation}
\frac{dy}{dx}=\frac{H\dot{a}-2a\dot{H}}{3H^{4}}
\end{equation}

where $\omega_{ch}$ is the EoS parameter for MCG determined as
\begin{equation}\label{10}
\omega_{ch}=\frac{p_{ch}}{\rho_{ch}}=A-\frac{B}{(3H^{2}u)^{\alpha+1}}
\end{equation}

\begin{equation}
\dot{H}=\frac{M^2}{18f_{0}\beta^{2}H^{2}-2M^{2}}\left[8\pi
G\left(3H^2(u(A+1)+v)-\frac{B}{(3H^{2}u)^{\alpha}}\right)-\frac{2k}{a^2}\right]
\end{equation}

Here $H$ is given by,
\begin{equation}
H=\sqrt{\frac{M^{2}}{9f_{0}\beta}-\frac{8GM^{2}\pi
u}{9f_{0}\beta}-\frac{8GM^{2}\pi
v}{9f_{0}\beta}+\frac{\sqrt{\left(-2a^{2}M^{2}+16a^{2}GM^{2}\pi
u+16a^{2}GM^{2}\pi
v\right)^{2}+72a^{2}f_{0}kM^{2}\beta}}{18a^{2}f_{0}\beta}}
\end{equation}

\subsection{Critical Points}

The critical points of the above system are obtained by putting
$\frac{du}{dx}=\frac{dv}{dx}=\frac{dy}{dx}=0$. But due to the
complexity of these equations, it is not possible to find a
solution in terms of all the involved parameters. \textbf{So we
are compelled to search for a viable numerical solution. In order
to accomplish this, we have to replace the parameters by suitable
numerical values which must be realistic cosmologically. As far as
the parameters of MCG are concerned, we will be using the values
which are consistent with the observational data \cite{Debnath,
Chakraborty, Ranjit}. We consider $A=1/3,~~ B=0.5,~~ \alpha=1$.
The choice of the Aether gravity parameters will not be so
straightforward, since we do not have cosmologically constrained
parametric values at our disposal. In \cite{Clifton2} Clifton et
al has shown that in order to realize a more exotic but viable
scenario, one should have $c_{1}=c_{3}=0$. From eqn.(23), we get
an idea of the value of $\beta$. If we take $\Omega_{m}=0.3$
(acceptable value from the latest observational data) and $n=1$,
we see that in order to make $f_{0}>0$, $\beta$ should be
restricted to negative values. Although we get an idea about the
sign of $\beta$ yet there is no information available in
literature about the magnitude. So keeping in mind the simplicity
of calculations, we choose $\beta=-1$. Using the accepted values
of parameters in eqn.(23) we obtain $f_{0}=1.5$ approximately.
Since $M$ has the dimensions of mass, so it is obvious that it
should have large values. The exact values of $M$ and $b$ are
obtained from numerical fine tuning after putting the accepted
numerical values of other parameters. We take, $M=5000,~~b=1$ and
obtain the following critical point,}
\begin{equation}
u_{c}=4.32\times10^{-14},~~~~~~ v_{c}=1.5,~~~~~~
~~~y_{c}=3.89729\times10^{-18}
\end{equation}

The critical point correspond to the era dominated by DM and MCG
type DE. For the critical point $(u_{c},v_{c})$, the equation of
state parameter given by equation (24) of the interacting DE takes
the form

\begin{equation}
\omega_{ch}=A-B\left[3u_{c}\left\{\frac{M^{2}}{9f_{0}\beta}-\frac{8GM^{2}\pi
u_{c}}{9f_{0}\beta}-\frac{8GM^{2}\pi
v_{c}}{9f_{0}\beta}+\frac{\sqrt{\left(-2a^{2}M^{2}+16a^{2}GM^{2}\pi
u_{c}+16a^{2}GM^{2}\pi
v_{c}\right)^{2}+72a^{2}f_{0}kM^{2}\beta}}{18a^{2}f_{0}\beta}\right\}\right]^{-1-\alpha}
\end{equation}

\vspace{2mm}

\begin{figure}
~~~~~~~~~~~~~~~~~~~~~~~~~~~~~~~~\includegraphics[scale=0.8]{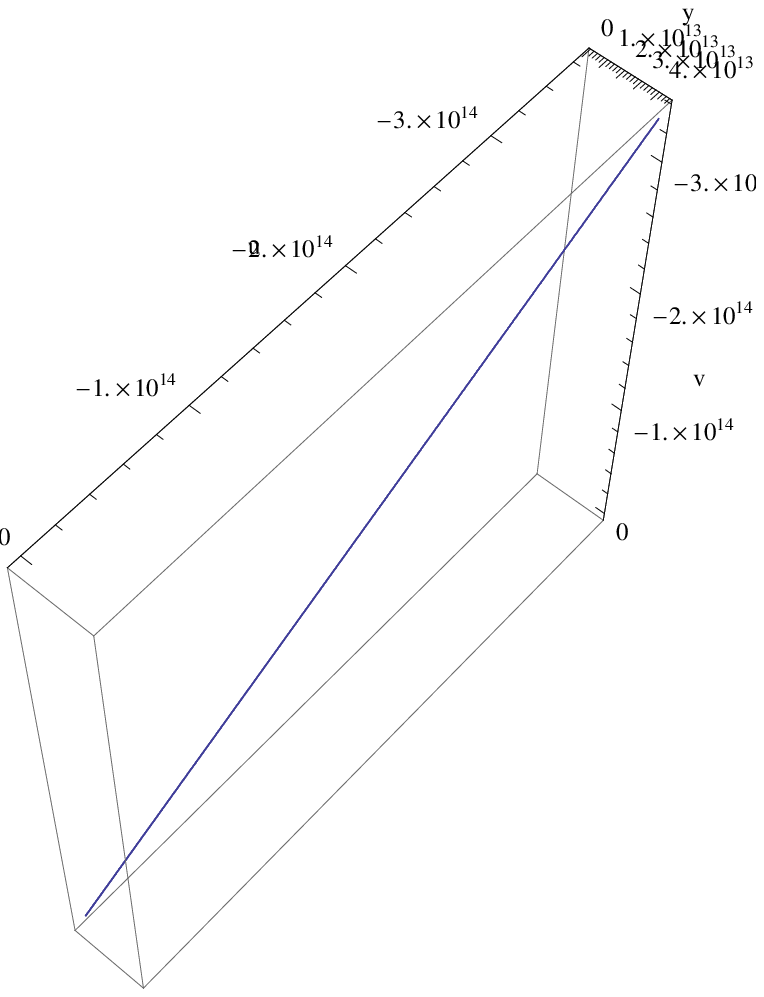}~~~~\\

~~~~~~~~~~~~~~~~~~~~~~~~~~~~~~~~~~~~~~~~~~~~~~~~~Fig.1~~~~~~~~~~~~~~~~~~~~~~~~~~~~~~~~~\\\\\\\\\\
\vspace{2mm} \textsl{Fig 1 : The dimensionless density parameters
$u$, $v$ and $y$ are plotted against each other in a 3D-scenario.
Other parameters are fixed at $\alpha=1,\beta=-1, b=1,
A=1/3, B=0.5, f_0=1.2$ and $M=5000$.}\\
\vspace{2mm}
\end{figure}

\begin{figure}
\includegraphics[scale=0.3]{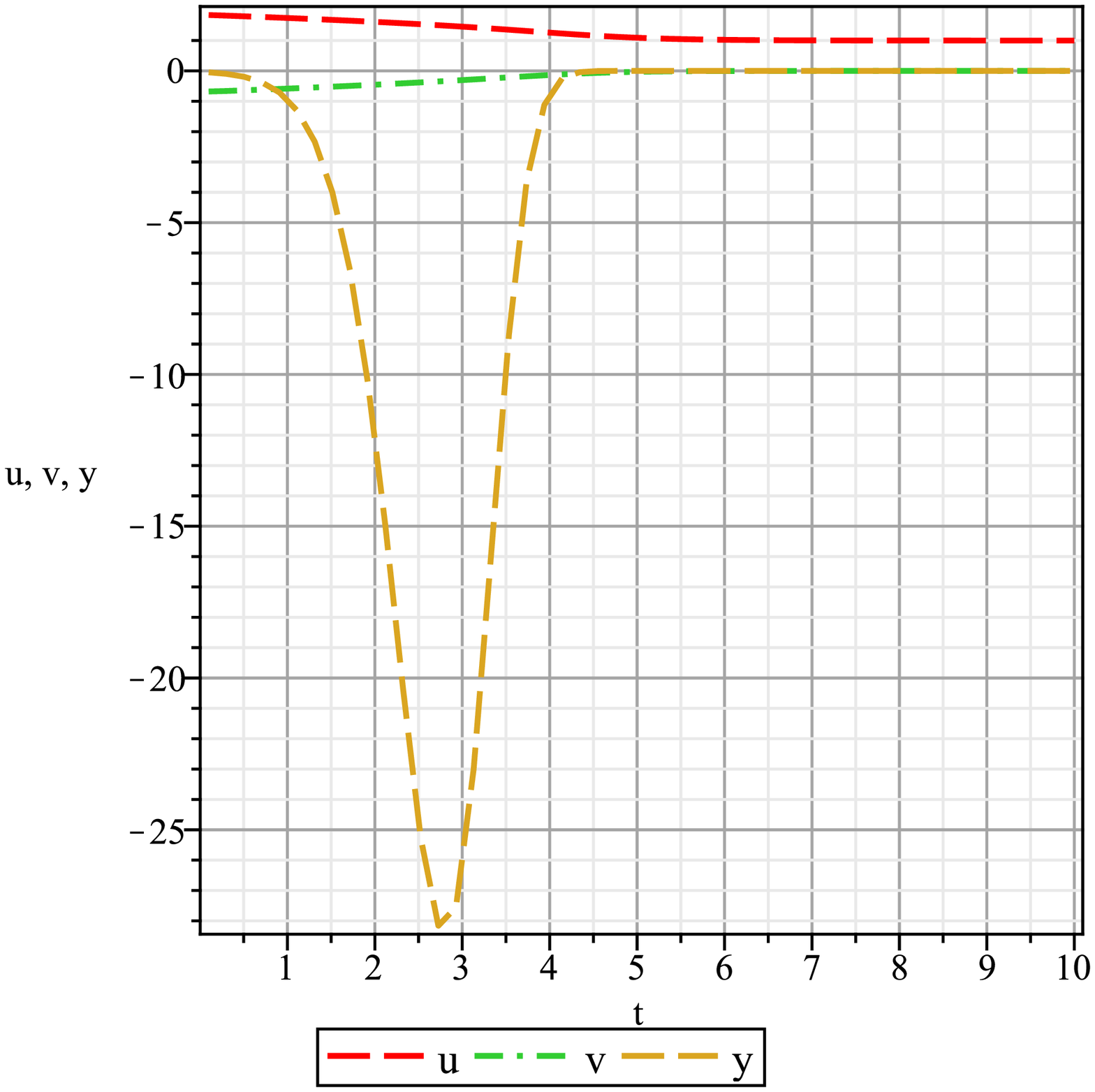}~~\includegraphics[scale=0.3]{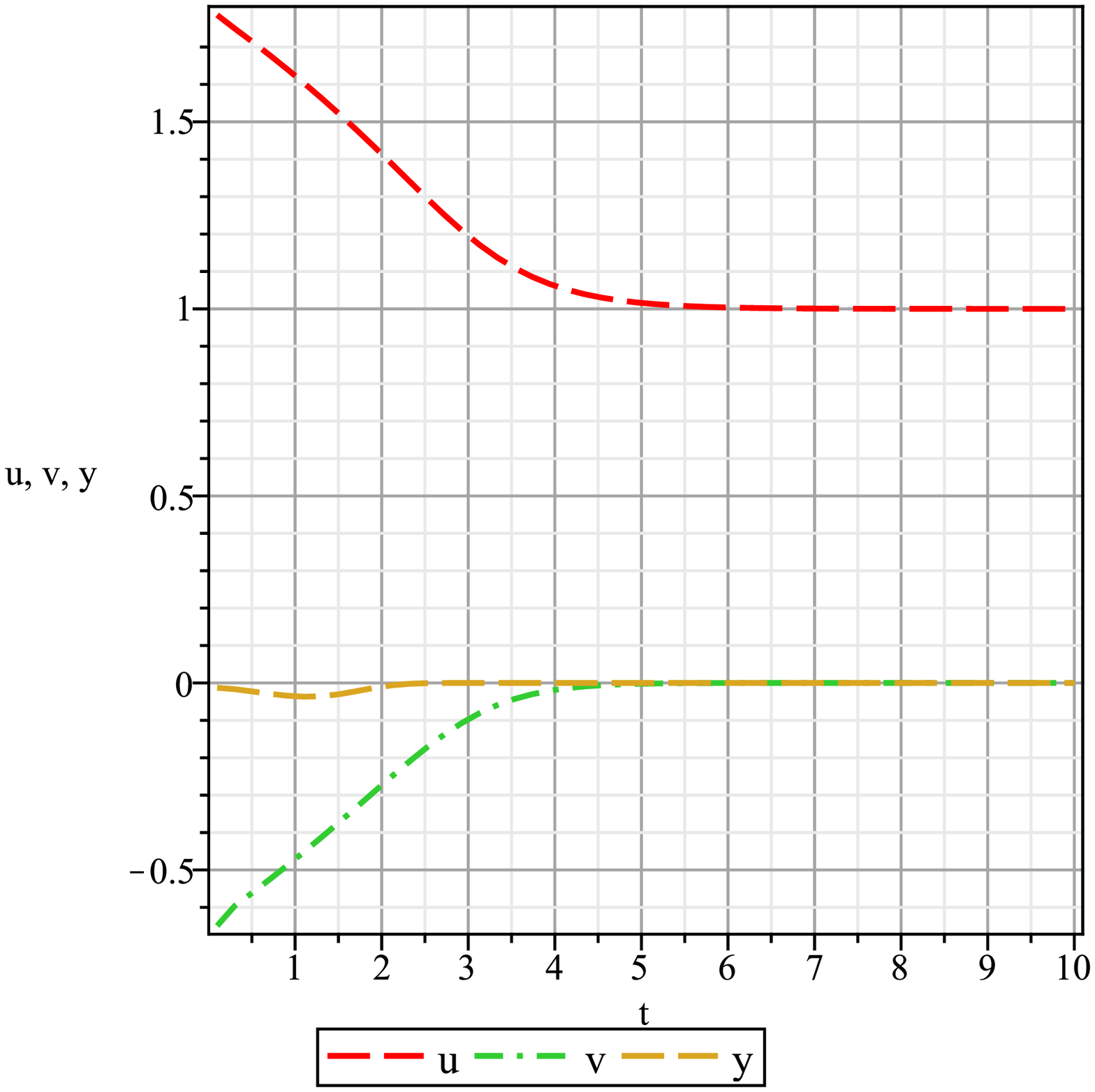}~~\includegraphics[scale=0.3]{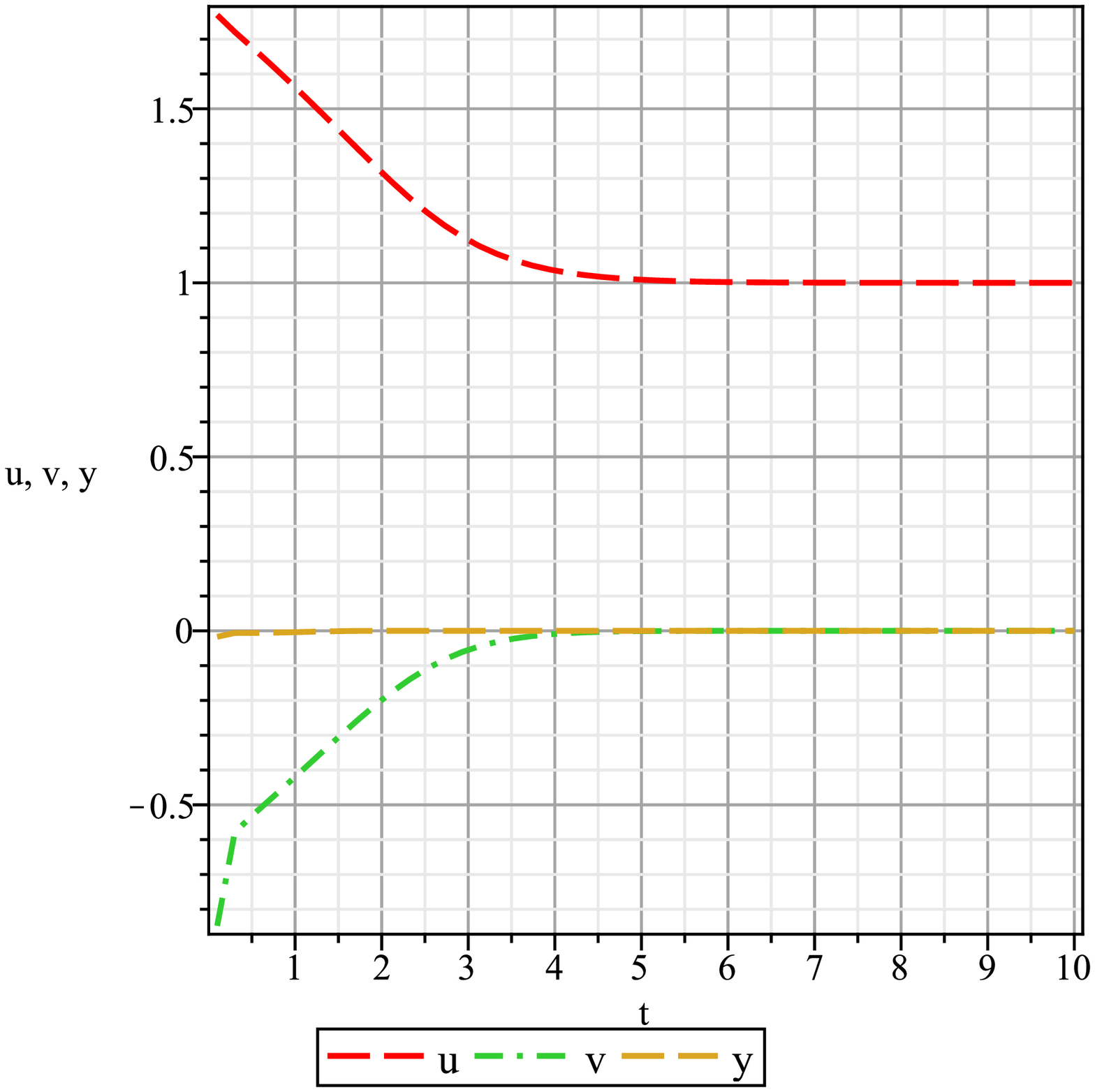}~

~~~~~~~~~~~~~~~~~~~~Fig.2~~~~~~~~~~~~~~~~~~~~~~~~~~~~~~~~~~~~~~~~~~~~~~~~~~~~~~~~~~~~~~Fig.3~~~~~~~~~~~~~~~~~~~~~~~~~~~~~~~~~~~~~~~~~~~~~~~~~~~~~~~~~~~~~~~~~~~Fig.4~~~~~~~~~~~~~~~~~~~~~~~~~~~~~~~~~\\\\\\\\
\vspace{1mm} \textsl{Fig 2 : The dimensionless density parameters
are plotted against e-folding time. The initial conditions are
$u(0.01)=2.5, v(0.01)=0.02$ and $y(0.01)=0.01$. Other parameters
are fixed at $\alpha=1,\beta=-1, b=0.3,
A=1/3, B=0.5, f_0=1.2$ and $M=2000$.\\
Fig 3 : The dimensionless density parameters are plotted against
e-folding time. The initial conditions are $u(0.01)=2.5,
v(0.01)=0.02$ and $y(0.01)=0.01$. Other parameters are fixed at
$\alpha=1,\beta=-1, b=0.4,
A=1/3, B=0.5, f_0=1.2$ and $M=2000$.\\
Fig 4 :  The dimensionless density parameters are plotted against
e-folding time. The initial conditions are $u(0.01)=2.5,
v(0.01)=0.02$ and $y(0.01)=0.01$. Other parameters are fixed at
$\alpha=1,\beta=-1, b=0.5,
A=1/3, B=0.5, f_0=1.2$ and $M=2000$.}\\
\end{figure}

\begin{figure}
\includegraphics[scale=0.3]{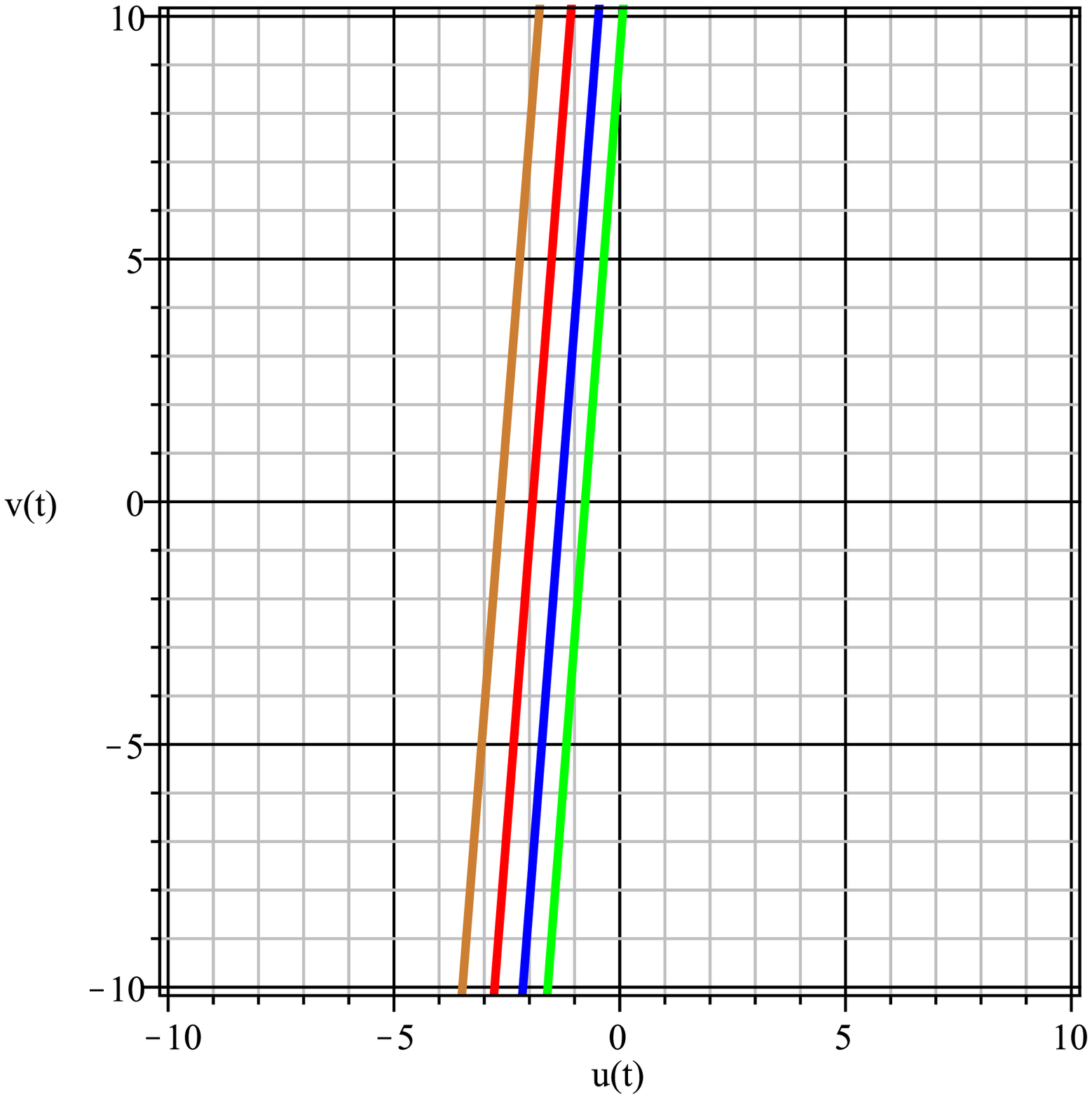}~~~\includegraphics[scale=0.3]{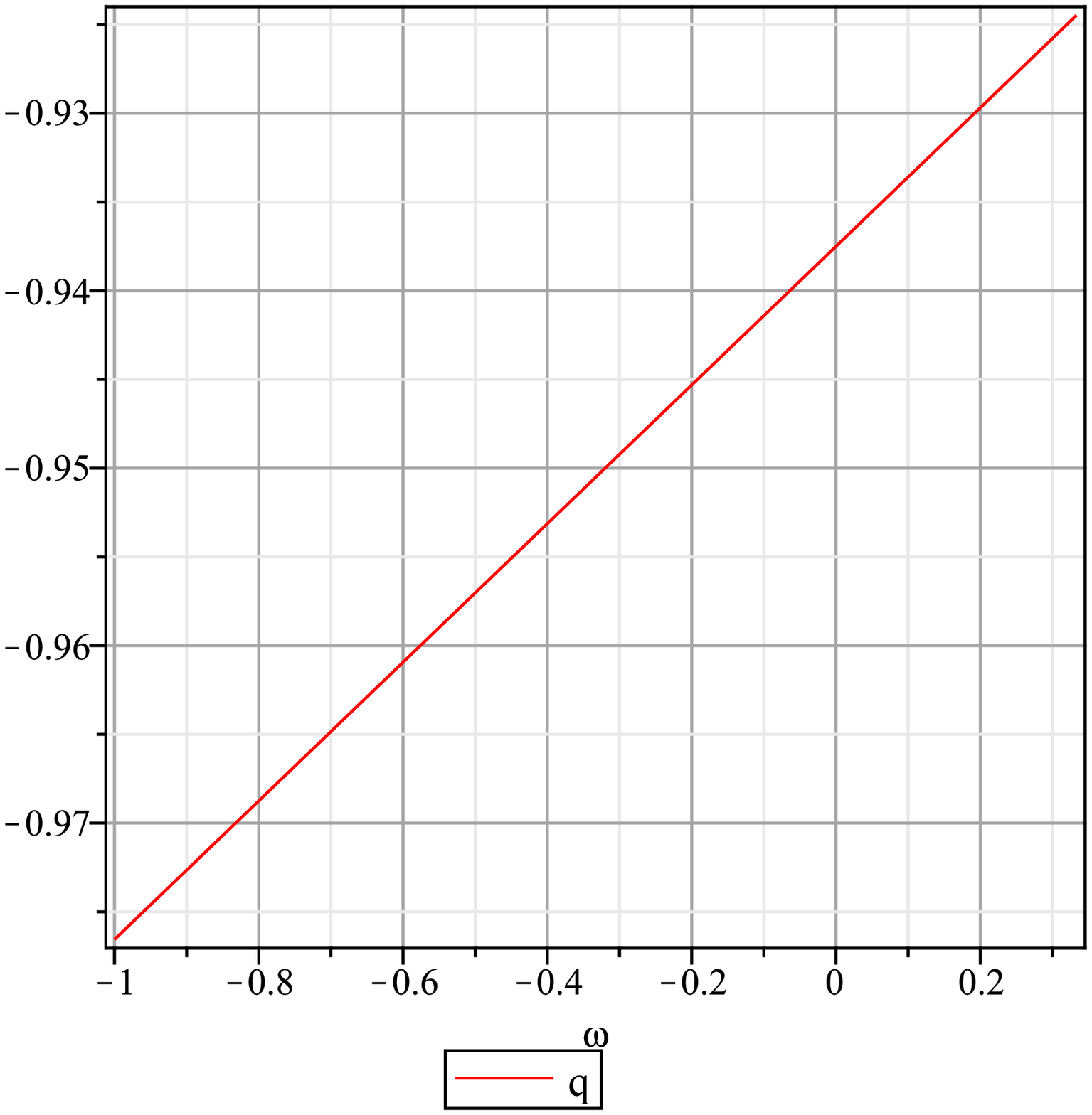}~~~\includegraphics[scale=0.3]{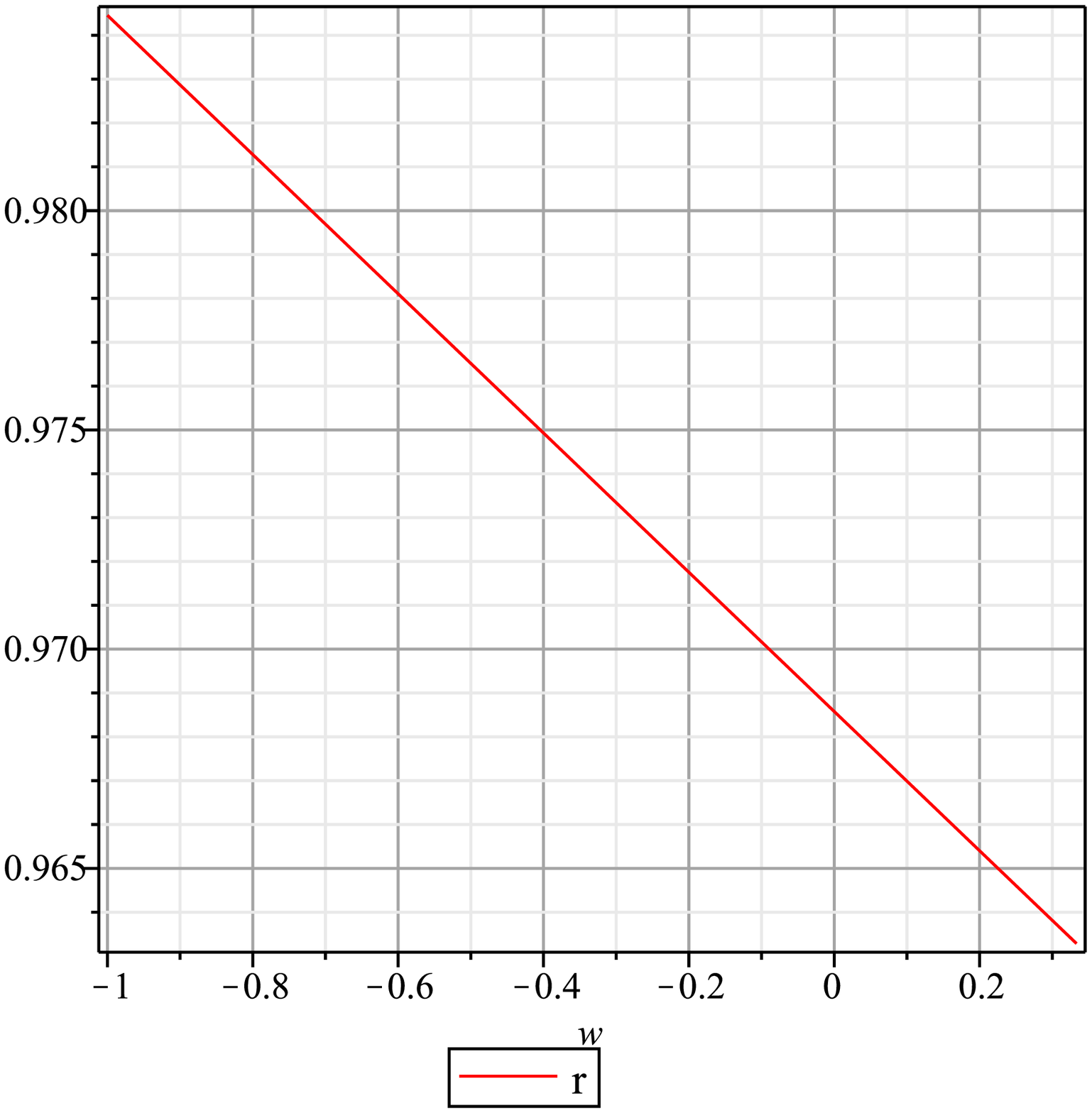}

~~~~~~~~~~~~~~Fig.5~~~~~~~~~~~~~~~~~~~~~~~~~~~~~~~~~~~~~~~~~~Fig.6~~~~~~~~~~~~~~~~~~~~~~~~~~~~~~~~~~~~~~~~~Fig.7~
\vspace{1cm}

\textsl{Fig 5: The phase diagram of the parameters $u(t)$ and
$v(t)$ depicting an attractor solution. The initial conditions
chosen are $u(1)=2.5, v(1)=0.05, y(1)=1.8$ (green); $u(1)=2.6,
v(1)=0.06, y(1)=1.9$ (blue); $u(1)=2.7, v(1)=0.07, y(1)=2.0$
(red); $u(1)=2.8, v(1)=0.08, y(1)=2.1$ (gold). Other parameters
are fixed at $\alpha=1,\beta=-1, b=1,
A=1/3, B=0.5, f_0=1.2$ and $M=2000$.\\
Fig 6 :   The deceleration parameter is plotted against the EoS
parameter. Other parameters are fixed at $\alpha=1,\beta=-1, b=1,
A=1/3, B=0.5, f_0=1.2$ and $M=2000$.\\
Fig 7 :  The statefinder parameter $r$ is plotted against the EoS
parameter. Other parameters are fixed at $\alpha=1,\beta=-1, b=1,
A=1/3, B=0.5, f_0=1.2$ and $M=2000$.}
\\

\end{figure}

\begin{figure}
\includegraphics[scale=0.3]{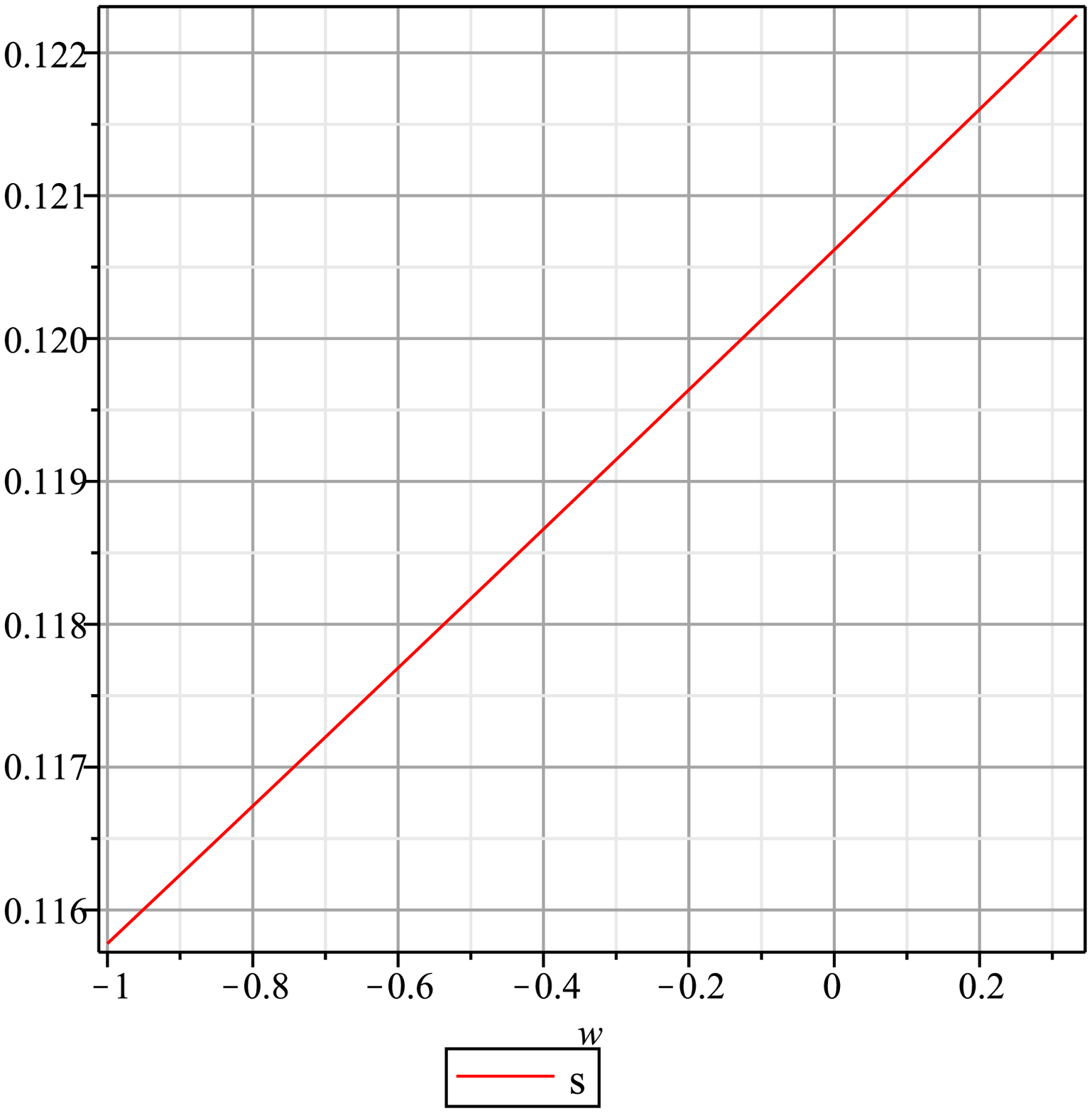}~~~~~~~~~~~~~\includegraphics[scale=0.3]{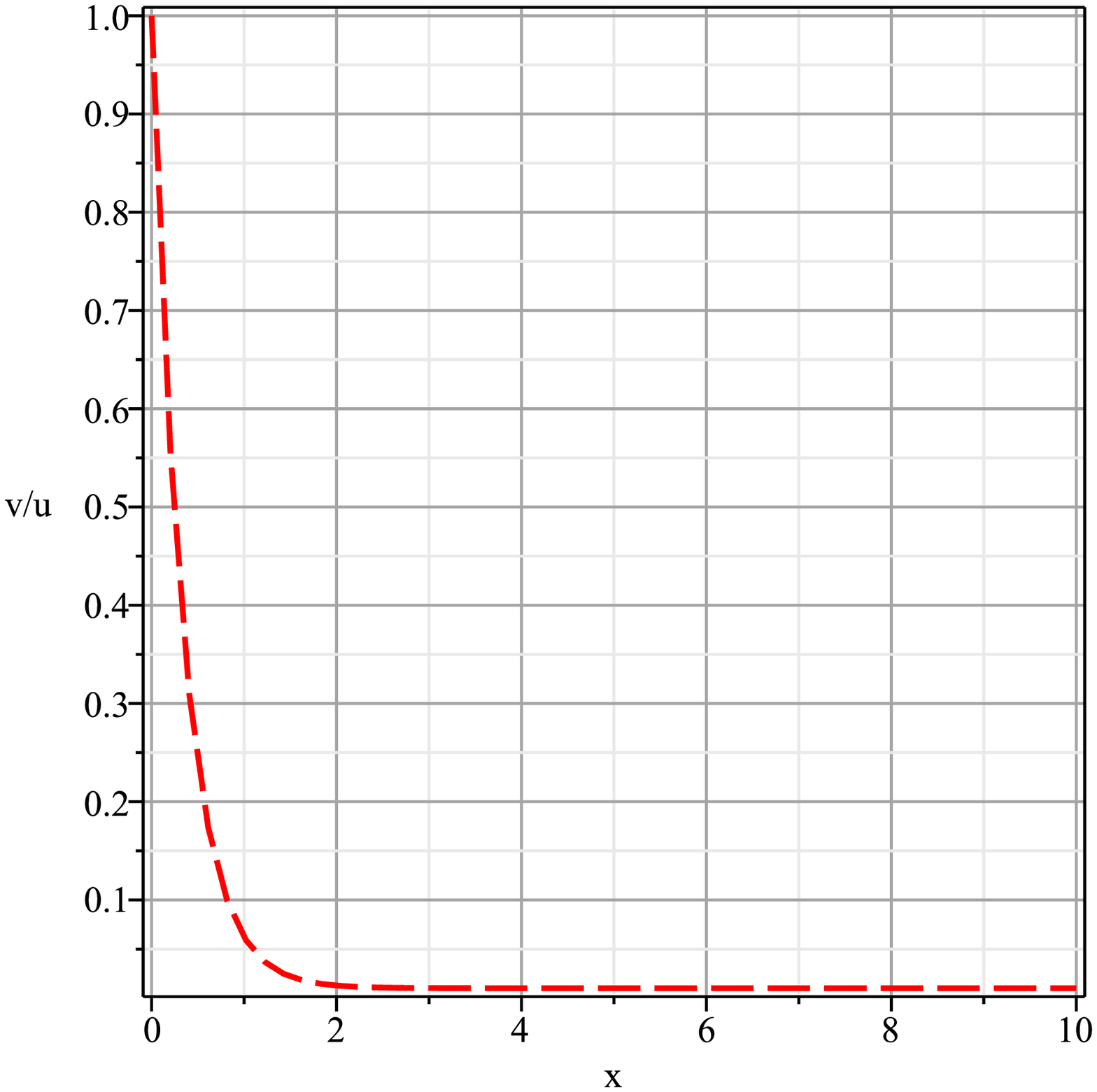}~~~~\\

~~~~~~~~~~~~~~~~~~~~~~~~~Fig.8~~~~~~~~~~~~~~~~~~~~~~~~~~~~~~~~~~~~~~~Fig.9\\
\vspace{1mm} \textsl{Fig 8 : The statefinder parameter $s$ is
plotted against the EoS parameter. Other parameters are fixed at
$\alpha=1,\beta=-1, b=1,
A=1/3, B=0.5, f_0=1.2$ and $M=2000$.\\
Fig 9 : The ratio of density parameters is shown against e-folding
time. The initial conditions chosen are v(1)=0.05, u(1)=2.5,
y(1)=1.8. Other parameters are fixed at $\alpha=1,\beta=-1, b=1,
A=1/3, B=0.5, f_0=1.2$ and $M=2000$.}
\\
\end{figure}

\subsection{Stability Around Critical Point}

Now we check the stability of the dynamical system  (eqs. (21) and
(22) and (23)) about the critical point. In order to do this, we
linearize the governing equations about the critical point i.e.,
\begin{equation}
u=u_c+\delta u ,~~~~   v=v_c+\delta v,~~~ ~~~ y=y_{c}+\delta y
\end{equation}

Now if we assume $f=\frac{du}{dx}$ , $g=\frac{dv}{dx}$ and
$h=\frac{dy}{dx}$ then we may obtain
\begin{equation}
\delta\left(\frac{du}{dx}\right)=\left[\partial_{u}
f\right]_{c}\delta u+\left[\partial_{v} f\right]_{c}\delta
v+\left[\partial_{y} f\right]_{c}\delta y
\end{equation}

\begin{equation}
\delta\left(\frac{dv}{dx}\right)=\left[\partial_{u}
g\right]_{c}\delta u+\left[\partial_{v} g\right]_{c}\delta
v+\left[\partial_{y} g\right]_{c}\delta y
\end{equation}
and

\begin{equation}
\delta\left(\frac{dy}{dx}\right)=\left[\partial_{u}
h\right]_{c}\delta u+\left[\partial_{v} h\right]_{c}\delta
v+\left[\partial_{y} h\right]_{c}\delta y
\end{equation}

\vspace{2mm}

where

\vspace{2mm}

\begin{eqnarray*}
\partial_{u}{f}=\frac{1}{3 f_0 u \beta }2^{1-\alpha } M^2 \left(-\frac{M^2 u (-1+u+v)}{f_{0}\beta}
\right)^{-\alpha }\left(3^{1+\alpha } B f_{0} \beta  (-u (-1+2
u+v) (-1+\alpha )+3 f_0 y (-u+(-1+2 u+v) \alpha ) \beta )\right.
\end{eqnarray*}
\begin{eqnarray*}
-2^{1+\alpha }f_0 \left(-\frac{M^2 u (-1+u+v)}{f_0 \beta
}\right)^{1+\alpha } \beta \left(4 (1+A) u^2+3 (1+A+b) f_0 y \beta
+u (-2-2 A+5 v+2 A v-9 (1+A+b)f_0 y \beta )\right.
\end{eqnarray*}
\begin{equation}
+v (-1+v-3 (1+A+3 b) f_0 y \beta )))
\end{equation}

\vspace{2mm}

\begin{eqnarray*}
\partial_{v}{f}=-2^{1-\alpha } 3^{\alpha } B M^2 (-1+\alpha ) \left(-\frac{M^2 u
(-1+u+v)}{f_{0} \beta }\right)^{-\alpha } (u-3 f_{0} y \beta
)~~~~~~~~~~~~~~~~~~~~~~~~~~~~~~~~~~~~~~~~~~~~~~~~~~~~~~~~~~~~~~~~~~~~~~~~~~~~~~~~
\end{eqnarray*}
\begin{equation}
+\frac{4 M^4 (-1+u+v) (u (-1+(3+2 A) u+3 v)-3 f_{0} (2 (1+A) u+b
(-1+3 u+3 v)) y \beta )}{3 f_{0}\beta }
\end{equation}

\vspace{2mm}

\begin{equation}
\partial_{y}{f}=2 M^2 (-1+u+v) \left(-2 M^2 (-1+u+v) ((1+A+b) u+b v)
-2^{-\alpha } 3^{1+\alpha } B f_0 \left(-\frac{M^2 u (-1+u+v)}
{f_0 \beta }\right)^{-\alpha } \beta \right)
\end{equation}

\vspace{2mm}

\begin{eqnarray*}
\partial_{u}{g}=-\frac{1}{u (1+2 (-1+u+v) \beta )^2}2^{1-\alpha }
M^2 y \left(-\frac{M^2 u (-1+u+v)}{f_0 \beta }\right)^{-\alpha }
~~~~~~~~~~~~~~~~~~~~~~~~~~~~~~~~~~~~~~~~~~~~~~~~~~~~
\end{eqnarray*}
\begin{eqnarray*}
\left(2^{1+\alpha } b u (-1+u+v) (-1+3 u+3 v) \left(-\frac{M^2 u
(-1+u+v)}{f_0 \beta }\right)^{\alpha } (M+2 M (-1+u+v) \beta
)^2+\right.~~~~~~~~~~~~~~~~~~~~~~~~~~~~~~~~~~~~~~~~~
\end{eqnarray*}
\begin{eqnarray*}
v \left(3^{1+\alpha } B f_0 \beta  (-u-\alpha +2 u \alpha +v
\alpha +2 (-1+u+v) (-1+2 u+v) \alpha  \beta )+\right.2^{1+\alpha }
f_0 \left(-\frac{M^2 u (-1+u+v)}{f_0 \beta }\right)^{1+\alpha }
\beta
\end{eqnarray*}
\begin{equation}
\left.\left.\left(-(-1+u+v) (-1+2 \beta ) (3+4 (-1+u+v) \beta )+A
\left(-1+v+4 u^2 \beta +2 (-1+v)^2 \beta +u (3+6 (-1+v) \beta
)\right)\right)\right)\right)
\end{equation}

\vspace{2mm}

\begin{equation}
\partial_{v}{g}=-\frac{1}{(1+2 (-1+u+v) \beta )^2}2^{1-\alpha } M^2 y
\left(-\frac{M^2 u (-1+u+v)}{f_0 \beta }\right)^{-\alpha }
~~~~~~~~~~~~~~~~~~~~~~~~~~~~~~~~~~~~~~~~~~~~~~~~~~~~~~~~~~~~~~~~~~~~~~~~~~~~~~~~~
\end{equation}
\begin{eqnarray*}
\left(-3^{1+\alpha } B f_0 \beta  (-1+u+2 v-v \alpha +2 (-1+u+v)
(-1+u+v-v \alpha ) \beta )+\right.
~~~~~~~~~~~~~~~~~~~~~~~~~~~~~~~~~~~~~~~~~~~~~~~~~~~~~~~~~~~~~~~~~~~~~~~~~~~~~~~
\end{eqnarray*}
\begin{eqnarray*}
2^{1+\alpha } M^2 (-1+u+v) \left(-\frac{M^2 u (-1+u+v)}{f_0 \beta
}\right)^{\alpha }\left(-1+2 u+A u-u^2-A u^2+5 v-5 u v-3 A u v-4
v^2-\right.
\end{eqnarray*}
\begin{eqnarray*}
2 (-1+u+v) \left((-1+u) (-2+u+A u)+2 (-4+(2+A) u) v+3 v^2\right)
\beta +4 (-1+u+v)^2 (-1+u+3 v) \beta ^2+
\end{eqnarray*}
\begin{equation}
\left.\left.b (-1+3 u+3 v) (1+2 (-1+u+v) \beta )^2\right)\right)
\end{equation}

\begin{eqnarray*}
\partial_{y}{g}=2 M^2 (-1+u+v)\left(-2 b M^2 (-1+u+v) (u+v)+\frac{1}
{1+2 (-1+u+v) \beta
}\right.~~~~~~~~~~~~~~~~~~~~~~~~~~~~~~~~~~~~~~~
\end{eqnarray*}
\begin{equation}
\times\left.v \left(2^{-\alpha } 3^{1+\alpha } B f_0
\left(-\frac{M^2 u (-1+u+v)}{f_0 \beta }\right)^{-\alpha } \beta
+2 M^2 (-1+u+v) (-1+u+A u+v-2 (-1+u+v) \beta )\right)\right)
\end{equation}

\vspace{2mm}

\begin{eqnarray*}
\partial_{u}{h}=\frac{1}{u (1+2 (-1+u+v) \beta )^2}2^{1-\alpha }M^2 y
\left(-\frac{M^2 u (-1+u+v)}{f_0 \beta }\right)^{-\alpha }
~~~~~~~~~~~~~~~~~~~~~~~~~~~~~~~~~~~~~~~~~~~~~~~~~~~~~~~~~~~~~~~~~~~
\end{eqnarray*}
\begin{eqnarray*}
\left(-3^{2+\alpha } B f_0 \beta  (-u-\alpha +2 u \alpha +v \alpha
+2 (-1+u+v) (-1+2 u+v) \alpha  \beta )-2^{1+\alpha } f_0
\left(-\frac{M^2 u (-1+u+v)}{f_0 \beta }\right)^{1+\alpha }
\beta\right.
\end{eqnarray*}
\begin{equation}
\left.\left(-1-3 A+9 u+9 A u+9 v+3 A v+2 (-1+u+v) (1+6 u+6 v+3 A
(-1+2 u+v)) \beta +8 (-1+u+v)^2 \beta ^2\right)\right)
\end{equation}

\vspace{2mm}

\begin{eqnarray*}
\partial_{v}{h}=\frac{1}{(1+2 (-1+u+v) \beta )^2}2^{1-\alpha } M^2 y
\left(-\frac{M^2 u (-1+u+v)}{f_0 \beta }\right)^{-\alpha }
~~~~~~~~~~~~~~~~~~~~~~~~~~~~~~~~~~~~~~~~~~~~~~~~~~~~~~~~~~~~~~~~~~
\end{eqnarray*}
\begin{eqnarray*}
\left(-3^{2+\alpha } B f_0 \beta  (-1+\alpha +2 (-1+u+v) \alpha
\beta )+2^{1+\alpha } M^2 (-1+u+v) \left(-\frac{M^2 u
(-1+u+v)}{f_0 \beta }\right)^{\alpha }\right.
\end{eqnarray*}
\begin{equation}
\left.\left(-1+9 u+6 A u+9 v+2 (-1+u+v) (1+3 (2+A) u+6 v) \beta +8
(-1+u+v)^2 \beta ^2\right)\right)
\end{equation}
\vspace{2mm}

\begin{eqnarray*}
\partial_{y}{h}=-\frac{1}{1+2 (-1+u+v) \beta }2 M^2 (1-u-v)
~~~~~~~~~~~~~~~~~~~~~~~~~~~~~~~~~~~~~~~~~~~~~~~~~~~~~~~~~~~~~~~~~~~~~~~~~~~~
\end{eqnarray*}
\begin{equation}
\left(2^{-\alpha } 3^{2+\alpha } B f_0 \left(-\frac{M^2 u
(-1+u+v)}{f_0 \beta }\right)^{-\alpha } \beta +2 M^2 (-1+u+v) (1+3
u+3 A u+3 v+2 (-1+u+v) \beta )\right)
\end{equation}

\vspace{2mm}

The Jacobian matrix of the above system is given by,
$$
J_{\left(u,v,y\right)}^{(EA)}=\left(\begin{array}{c}\frac{\delta
f}{\delta u} ~~~~~ \frac{\delta f}{\delta v} ~~~~~ \frac{\delta
f}{\delta y}\\ \frac{\delta g}{\delta u}~~~~~ \frac{\delta
g}{\delta v} ~~~~~ \frac{\delta g}{\delta y}\\ \frac{\delta
h}{\delta u} ~~~~~ \frac{\delta h}{\delta v} ~~~~~ \frac{\delta
h}{\delta y}\end{array}\right)
$$

The eigen values of the above matrix are calculated at the
critical point $(u_{c}, v_{c}, y_{c})$ and are found to be~~ \\

${\bf \lambda_{1}=1.40623\times 10^{15},~~~
\lambda_{2}=1.56236\times 10^{14}, ~~~
\lambda_{3}=-0.00543761}$.\\

Hence it is a {\bf Saddle Point}.

\vspace{2mm}

\subsection{Nature of cosmological parameters}
In EA model, the deceleration parameter $q$ can be obtained as
\begin{eqnarray*}
q^{(EA)}=-1-\frac{3}{2}
~~~~~~~~~~~~~~~~~~~~~~~~~~~~~~~~~~~~~~~~~~~~~~~~~~~~~~~~~~~~~~~~~~~~~~~~~~~~~~~~~~~~~~~~~~~~~~~~~~~~~~~~~~~~~~
\end{eqnarray*}
\begin{equation}
\times\frac{\left\{2M^2\left(-\frac{2k}{a^2}+8G\pi(p+\rho
)\right)\right\}}{\left[3\left\{\frac{M^2}{9f_{0} \beta
}+\frac{\sqrt{a^4 M^4+18 a^2f_{0} k M^2 \beta +48 a^4f_{0} G M^2
\pi\beta\rho}}{9 a^2f_{0} \beta}\right\} \left\{-2M^2+18f_{0}
\beta ^2 \left(\frac{M^2}{9f_{0} \beta}+\frac{\sqrt{a^4 M^4+18
a^2f_{0} k M^2 \beta +48 a^4 f_{0} G M^2 \pi \beta \rho }}{9 a^2
f_{0} \beta}\right)\right\}\right]}
\end{equation}

From above we get,
\begin{equation}
q_c^{(EA)}=-1+\frac{3}{2}X_{(EA)}
\end{equation}
~~~ where ~~~~
\begin{eqnarray*}
X_{(EA)}=
~~~~~~~~~~~~~~~~~~~~~~~~~~~~~~~~~~~~~~~~~~~~~~~~~~~~~~~~~~~~~~~~~~~~~~~~~~~~~~~~~~~~~~~~~~~~~~~~~~~~~~~~~~~~~~
\end{eqnarray*}
\begin{equation}
-\frac{\left\{2M^2\left(-\frac{2k}{a^2}+8G\pi(p+\rho
)\right)\right\}}{\left[3\left\{\frac{M^2}{9f_{0} \beta
}+\frac{\sqrt{a^4 M^4+18 a^2f_{0} k M^2 \beta +48 a^4f_{0} G M^2
\pi\beta\rho}}{9 a^2f_{0} \beta}\right\} \left\{-2M^2+18f_{0}
\beta ^2 \left(\frac{M^2}{9f_{0} \beta}+\frac{\sqrt{a^4 M^4+18
a^2f_{0} k M^2 \beta +48 a^4 f_{0} G M^2 \pi \beta \rho }}{9 a^2
f_{0} \beta}\right)\right\}\right]}
\end{equation}
If ~~$\frac{2k}{a^{2}}=8\pi G\left(p+\rho\right),~~ ~~X_{(EA)}=0,
~~~$ we have $q=-1$, which confirms the accelerated expansion  of
the universe. When $M^{2}=9f_{0} \beta ^2 \left(\frac{M^2}{9f_{0}
\beta}+\frac{\sqrt{a^4 M^4+18 a^2f_{0} k M^2 \beta +48 a^4 f_{0} G
M^2 \pi \beta \rho }}{9 a^2 f_{0} \beta}\right)$ we have
$q=-\infty$. Therefore we have
super accelerated expansion of the universe.\\

In this scenario, the Hubble parameter can be obtained as,
\begin{equation}
H=\frac{2}{3X_{(EA)}t}
\end{equation}
where the integration constant has been ignored. Integration of
eqn.(44) yields
\begin{equation}
a(t)=a_0t^{\frac{2}{3X_{(EA)}}}
\end{equation}
which gives the power law form of expansion of the universe. In
order to have an accelerated expansion of universe in EA gravity
we must have $0<X_{(EA)}<\frac{2}{3}$. Using this range of
$X_{(EA)}$ in the equation $q_{c}^{EA}=-1+\frac{3}{2}X_{(EA)}$, we
get the range of $q_{c}^{(EA)}$ as $-1<q_{c}^{(EA)}<0$. This is
perfectly consistent with an accelerated expansion of the
universe.

Sahni et al (2003) \cite{Sahni} introduced a pair of cosmological
diagnostic pair $\{r,s\}$ which is known as as statefinder
parameters. The two parameters are dimensionless and are
geometrical since they are derived from the cosmic scale factor
alone. Also this pair generalizes the well-known geometrical
parameters like the Hubble parameter and the deceleration
parameter. The statefinder parameters are given by
\begin{equation}\label{27}
r\equiv\frac{\stackrel{...}a}{aH^3},\ \
s\equiv\frac{r-1}{3(q-1/2)}.
\end{equation}
In EA model, we have the following expressions of $r$ and $s$ as
\begin{equation}\label{rsbasic19}
r_{(EA)}=\left(1-\frac{3X_{(EA)}}{2}\right)\left(1-3X_{(EA)}\right).
\end{equation}
and
\begin{equation}\label{rsbasic20}
s_{(EA)}=X_{(EA)}
\end{equation}

\section{Graphical Analysis}

Graphs are obtained and phase diagrams are drawn in order to
determine the type of critical point obtained in this model. Below
we discuss the results obtained in detail:

The dimensionless density parameters $u$ and $v$ and $y$ are
plotted against each other in figure 1. In figs. 2,3 and 4, the
density parameters are plotted against time for different values
of interacting coupling parameter. From the figures we see that
$u$ always dominates over $v$ during evolution of the universe. So
this result is consistent with the well known idea of an energy
dominated universe.

Fig. 5 shows the phase portrait of the density parameters of DE
and DM. As already stated before that the critical point obtained
in this system is a Saddle point and hence there always remains a
question on the stability of the system. In fig.6, a plot of
deceleration parameter, $q$ is obtained against the EoS parameter,
$\omega$. It is seen that $q$ remains in the negative level thus
confirming the recent cosmic acceleration. Figs. 7 and 8 show the
plot of the statefinder parameters $r$ and $s$ respectively
against the EoS parameter, $\omega$. It is known that in case of
$\Lambda$CDM model $r=1$ and $s=0$. comparing figs. 7 and 8, it
can be seen that the values of $r$ is quite different from $1$
corresponding to the values of $\omega$ when $s=0$.

This gives the deviation of the model from the $\Lambda$CDM model.
Finally in fig.9, the ratio $v/u$ is plotted against $x=\ln a$.
The decreasing trajectory confirms the existence of an energy
dominated universe with progressive values of scale factor.

\section{Discussions and Concluding Remarks}
In this work, we have considered a combination of Modified
Chaplygin gas in Einstein-Aether gravity model. Our basic idea was
to study the background dynamics of MCG in detail when it is
incorporated in Aether gravity. Dynamical system analysis was
carried out, critical points were found and the stability of the
system around those critical points was tested. Graphical analysis
was done to get an explicit picture of the outcome of the work. In
order to find a solution for the cosmic coincidence problem, a
suitable interaction between DE and DM was considered. The set of
equations characterizing the dynamical system was formed and a
stable scaling solution was obtained. Hence this work can be
considered to be a significant one as far as solution of cosmic
coincidence problem is concerned. From the above analysis we
conclude that the combination of MCG in EA gravity makes a perfect
model for the expanding universe undergoing a late acceleration. \\

\section*{Acknowledgements}

The authors sincerely acknowledge the facilities provided by the
Inter-University Centre for Astronomy and Astrophysics (IUCAA),
pune, India where a part of the work was carried out. The authors
also thank the anonymous referee for his or her constructive
comments that helped them to improve the quality of the manuscript.\\

\end{document}